\documentclass[a4paper,fleqn,usenatbib]{mnras}

\usepackage{mathptmx}

\usepackage[T1]{fontenc}
\usepackage{ae,aecompl}

\usepackage{graphicx}	
\usepackage{amsmath}	
\usepackage{amssymb}

\newcommand{\lcdm}{$\Lambda$CDM}
\newcommand{\DoF}{degrees of freedom}
\newcommand{\ket}[1]{$| {#1} \rangle$}

\title{ Tuning Goodness-of-Fit Tests}
\author[Arrasmith et. al.]
  {A.~Arrasmith,$^1$\thanks{aarrasmith@ucdavis.edu}
  B.~Follin,$^1$ E. ~Anderes,$^2$ L.~Knox$^1$\\
  $^1$Department of Physics, University of California, Davis\\
  One Shields Ave, Davis, CA 95616\\
  $^2$Department of Statistics, University of California, Davis\\
  One Shields Ave, Davis, CA 95616}
\date{May 2017}

\begin{document}
\maketitle

\begin{abstract}
As modern precision cosmological measurements continue to show agreement with the broad features of the standard $\Lambda$-Cold Dark Matter (\lcdm) cosmological model, we are increasingly motivated to look for small departures from the standard model's predictions which might not be detected with standard approaches.
While searches for extensions and modifications of \lcdm{} have to date turned up no convincing evidence of beyond-the-standard-model cosmology, the list of models compared against \lcdm{} is by no means complete and is often governed by readily-coded modifications to standard Boltzmann codes. Also, standard goodness-of-fit methods such as a naive $\chi^2$ test fail to put strong pressure on the null \lcdm{} hypothesis, since modern datasets have orders of magnitudes more degrees of freedom than \lcdm{}.
Here we present a method of tuning goodness-of-fit tests to detect potential sub-dominant extra-\lcdm{} signals present in the data through compressing observations in a way that maximizes extra-\lcdm{} signal variation over noise and \lcdm{} variation. This method, based on a Karhunen-Lo\`{e}ve transformation of the data, is tuned to be maximally sensitive to particular types of variations characteristic of the tuning model; but, unlike direct model comparison, the test is also sensitive to features that only partially mimic the tuning model.
As an example of its use, we apply this method in the context of a nonstandard primordial power spectrum compared against the $2015$ $Planck$ CMB temperature and polarization power spectrum. We find weak evidence of extra-\lcdm{} physics, conceivably due to known systematics in the 2015 Planck polarization release.

\end{abstract}


\section{Introduction}
\label{sec:intro}

Perhaps the most important high-level cosmological result of the $Planck$ mission is the continuing consistency of CMB data with the standard cosmological model--a consistency that persists after a substantial increase in sensitivity and angular resolution for both temperature and polarization anisotropies \citep{planck_collaboration_planck_2016-2}. Given the importance of this result, we are motivated to consider the strengths and weaknesses of various consistency tests, and how they can be ``tuned'' to be maximally effective at testing for departures from the standard cosmological model. 

To date the tests of the consistency of the $Planck$ data are of three kinds: 1) searching for evidence for particular extensions of the standard cosmological model \citep[e.g.][]{planck_collaboration_planck_2015-1}, 2) tests for internal consistency of cosmological parameters derived from different subsets of the data \citep[e.g.][]{planck_collaboration_planck_2016}, and 3) ``goodness-of-fit'' tests via a $\chi^2$ statistic with a very large number of degrees of freedom \citep{planck_collaboration_planck_2015}. These  types of searches span extremes of a continuum. On one side of the extreme is the first of these three kinds of tests. These tests are the most powerful at finding particular departures from the standard cosmological model, as long as that departure is the one under consideration. On the other side of the extreme is the $\chi^2$ statistic with a very large number of degrees of freedom ($\sim 1$ for every measured multipole), which, for any particular type of departure, is much weaker than a direct search for that departure.

The $\chi^2$ test has, in principle, the advantage of sensitivity to a wide range of departures from \lcdm, but due to the large number of degrees of freedom in the data a failure of the $\chi^2$ test, when compared to a direct search for a particular extension, requires a much larger departure from \lcdm{} to fail.
By preserving the large number of degrees of freedom, a $\chi^2$ test applied to each multipole moment is sensitive to a much broader range of possible departures of the CMB power spectra than the set reachable via changes to cosmological parameters. For example, a big spike at $\ell = 273$, with no changes at other multipoles, large enough to cause a failure of the $\chi^2$ test, could remain undetected by a specific search for the very smooth departure to the angular power spectrum caused by running of the spectral index.

On the other hand, the signal from running with sufficient amplitude that it can be detected at $3\sigma$ or greater via parameter estimation in the $\Lambda$CDM + running parameter space would have to be $\simeq 4$ times bigger to cause a failure of the $\chi^2$ test on the binned $Planck$ high-$\ell$ temperature power spectrum. This specific example is part of a general result: the more specific the searches, the greater their sensitivity to the models they are designed to look for.

An ideal goodness-of-fit test is one that preserves sensitivity to a large range of physically possible departures from the null hypothesis, and keeps that sensitivity as high as possible by eliminating degrees of freedom that fail to show significant excitations under physically realistic signal variation. As we will see with explicit examples below, turning this vague idea into a precise prescription requires a precise description of our prior expectation on departures from the null hypothesis. Thus there is no single objectively optimal goodness-of-fit test. One can only optimize (or ``tune'') a test with the adoption of an alternative model space, including specific prior probabilities for the parameters of this space.

One can directly see the downside of a $\chi^2$ test with a large number of degrees of freedom $m$ following from the fact that the standard deviation in the expected value of $\chi^2$ increases as $\sqrt{m}$. A change to the signal model that increases the likelihood by a factor of 10,000, causes a change in $\chi^2$ by $\Delta \chi^2 = -2\ln{10^{-4}}= 18.4$. This is above $\sigma(\chi^2)$ for the case of, say, $m=40$ degrees of freedom, but well below it for the case of $m=3,000$ degrees of freedom.

In the 3,000 degrees of freedom case, the expected variation in the $\chi^2$ due to noise is large enough that the test is not sensitive to many
possible signal variations. As such, a $\chi^2$ test on this large number of degrees of freedom is primarily a test of the noise model rather than the signal model.

Recently, consistency tests of the $Planck$ temperature power spectrum data \citep{planck_collaboration_planck_2016,addison} have checked for consistency between cosmological parameter estimates from two different $\ell$ ranges. We can understand 
such an approach as a particular means of increasing the power of the test by greatly decreasing the degrees of freedom. Here we explore the tuning of goodness-of-fit tests in a manner that reveals the general principles driving the tuning.

To produce a $\chi^2$ test that is a powerful test of the signal model we want to perform a compression of the data that reduces the degrees of freedom. We want to do so in a manner that preserves exotic signal variation and reduces noise variation. With these statical properties specified, a linear transformation on the original degrees of freedom can be performed that orders these new modes according to their (exotic) signal-to-noise ratios.

The structure of this paper is as follows. In section \ref{sec:xc} we discuss $\chi^2$ and reduced-degree $\chi^2$ goodness-of-fit tests in cosmology. In section \ref{sec:methods} we introduce a method for finding good degrees of freedom on which to perform reduced-degree $\chi^2$ tests when considering a specific type of model alternative. In section \ref{sec:uses} we relate this technique to similar techniques that have been used in other roles in cosmology and show an example application to looking for departures from the \lcdm{} primordial power spectrum.

\section{A Reformulation of Goodness of Fit}
\label{sec:xc}
A $\chi^2$ statistic measures the discrepancy between the observed data and the estimated signal under a parametric model for the signal. We focus on the situation where the data are represented as a vector composed of some unknown signal vector $|s\rangle$ and a noise vector $|n \rangle$
\begin{equation}
\label{eq:null model}
|d\rangle = |s\rangle + |n \rangle;
\end{equation}
 with the noise $|n \rangle$ drawn from a nearly Gaussian distribution with mean zero and covariance matrix $N$.
Assuming the true but unknown signal is generated from a parametric model $|s(\theta)\rangle$  (in the case of cosmology, \lcdm)
the usual $\chi^2$ statistic is given by
\begin{equation}
\label{eq:chi2}
p =  \langle d  - {s}(\hat\theta)| N^{-1} | d - {s}(\hat\theta)\rangle,
\end{equation}
where $\hat\theta$ is an estimate of $\theta$ from the data $|d\rangle$.

If $|s(\theta)\rangle$ is linear in $\theta$, $N$ is full rank and $\hat\theta$ denotes the maximum likelihood estimate, then the estimated signal $|s(\hat\theta)\rangle$ represents a projection of $|d\rangle $ onto the linear space spanned by $|s(\theta)\rangle$, relative to the inner product $\langle u |N^{-1}| v\rangle$. This implies that $p$ follows a $\chi^2$ distribution with $m$ degrees of freedom:
\begin{align}
\label{eq:chi2v2}
& p \sim \chi^2_m (x)\\ 
&\langle p \rangle = m \nonumber \\ 
&\langle p^2 \rangle = 2m \nonumber,
\end{align}
where $m = k - r$ is the effective number of degrees of freedom, $k$ is the number of measurements and $r$ is the number of free parameters in the model (see \citet{Wichura_2006} or \citet{kruskal1961} for an excellent coordinate-free derivation).
If $| s(\theta)\rangle$ is nonlinear in $\theta$ then  equation \ref{eq:chi2v2} serves as an approximation when the data are informative enough to ensure that the difference between the estimate and truth, $\hat\theta - \theta_0$, is sufficiently small to allow a linear approximation of $|s(\theta)\rangle - |s(\hat\theta)\rangle$ with respect to $\hat\theta - \theta_0$. Under this approximation we can define the PTE ($p$-value) of $p$, or probability to exceed $p$ in the distribution, as
\begin{equation}
\label{eq:PTE}
PTE = \int_p^{\infty} \chi^2_{m}(x) dx,
\end{equation}
which is the probability of $p$ or a more extreme value occurring under the assumption of the null \lcdm{} model (and noise).

A low PTE reflects a rare (unlikely) event under the null model, and is thus interpreted as evidence for the presence of fluctuations captured neither by the noise covariance nor the null model. In other words, the data vector in equation \ref{eq:null model} has an additional, extra-\lcdm{} component $|s_\Xi\rangle$ not captured by the null model or the noise. The two primary benefits of this approach over, say, simply fitting for the presence of $|s_\Xi \rangle$ are: 1) the $\chi^2$ test can be much more computationally efficient than fitting, especially if the $|s_\Xi \rangle$ is costly to compute or has many additional parameters to estimate and 2) that 
the $\chi^2$ test trades the ability to estimate the presence of a particular feature in the data for sensitivity to a more general set of features (any feature that correlates with extra-\lcdm{} signal).

In equation \ref{eq:chi2}, the covariance $N$ is taken to be the covariance of all sources of variation of the data. This covariance is \emph{a priori} unknown, and is usually (in frequentist literature) taken to be the variance in the noise contribution 
\begin{equation}
|n\rangle \sim \mathcal{N}(0, N).
\end{equation}
Even under the assumption of the null model, for $N$ to capture all variation in $|d\rangle$ in equation \ref{eq:chi2}, it must be that
\begin{equation}
|s \rangle \simeq |s(\hat\theta)\rangle
\end{equation}
for the best-fit $|s(\hat\theta)\rangle$. A useful metric for the expected deviation between the truth and the best-fit model under the null model is variation in the posterior distribution of the null model under the data. If the variation in the data caused by changing \lcdm{} parameters is smaller compared to fluctuation due to the noise, then $N$ is an accurate stand-in for the total covariance of the data.
 An analogous point applies to the interpretation of $p$ as a test statistic under equation \ref{eq:PTE}. For a small PTE to be interpreted as evidence for an additional component $|s_\Xi\rangle$, the size of this signal must be comparable to or greater than the expected fluctuations in the noise. If this criterion is not satisfied for a given model of the fluctuations $|{s_\Xi}\rangle$, the $\chi^2$ test statistic is not sensitive to that model.

In the context of CMB experiments, correlations over multipoles $\ell$ in the \lcdm{}
arise generically since projection effects and gravitational lensing at recombination and along the line of sight smooth out sharp features in the CMB power spectra \citep{hu_principal_2004}. This smoothness basically means that any cosmological model that is based on the same general framework as \lcdm{} will excite significantly fewer modes in the data than are measured. Since CMB measurements have so many \DoF{}, each of which inherently has some noise, the expected variations in the $\chi^2$ test statistic from the noise are large enough to mask any signal \ket{s_\Xi} that isn't particularly large. 
Given that \lcdm{} fits the data so well, we don't expect to find many large signals from extra-\lcdm{} models and therefore the test in equation \ref{eq:PTE} is, in its unadulterated form, insensitive to most plausible extensions to \lcdm{}. 


To tackle the problem of $\chi^2$ insensitivity, we begin by introducing a change of basis on the measured \DoF{} in equation \ref{eq:chi2}. We introduce a set of filters (vectors) $|v_i\rangle$ that span the space defined by the \DoF{} inherent in the data:
\begin{equation}
|v_i\rangle \in \mathbb{R}^k, \,\, i\in \left\{1,...,k\right\},
\end{equation}
where $k$ is the number of \DoF{} in the data. Equation \ref{eq:chi2} can be re-written as
\begin{equation}
\label{eq:chi2_2}
p = \sum_{i,j} \langle v_i | d - {s}(\hat\theta)\rangle N^{-1}_{ij} \langle d - {s}(\hat\theta) | v_j \rangle ,
\end{equation}
where $N_{ij} = \langle v_i | N | v_j \rangle$ is the new noise covariance in the basis $|v_i\rangle$.

Equation \ref{eq:chi2_2} is equivalent to \ref{eq:chi2}. However, for a judicious choice of the $|v_i \rangle$--namely, a choice where at least some $|v_i \rangle$ have large-scale correlations across multipoles--it is possible for signal variation in the exotic model to become comparable to the noise fluctuation for a given mode (or subset of modes) $|v_j\rangle$. To reinstate goodness-of-fit as a test of the model, what we then seek is a basis $|v_i\rangle$ where the exotic signal variation projected along some subset $|v_j\rangle$ of these modes is greater than, or at least comparable to, the projected fluctuations of the noise:
\begin{equation}
\label{eq: signal-to-noise}
\langle v_j| \Sigma_\Xi | v_j \rangle \gtrsim \langle v_j | N |v_j \rangle,
\end{equation}
where $\Sigma_\Xi$ is the covariance of some prior distribution of extra-\lcdm{} fluctuations. If we restrict the sum in equation \ref{eq:chi2_2} under such a set $|v_i\rangle$ to the set of modes
\begin{equation}
S = \left\{|v_j\rangle | \langle v_j| \Sigma_\Xi | v_j \rangle \gtrsim \langle v_j | N |v_j \rangle\right\} ,
\end{equation}
where the fluctuations in exotic signal are non-negligible compared to the noise in that mode, we enhance the ability of the statistic to detect insufficiencies in the null \lcdm{} model. Note that restricting the sum to this set and excluding the other degrees of freedom, not simply changing the basis, is what increases the sensitivity of the statistic. The problem reduces to finding the elements of the set $S$, with one important caveat: since the degrees of freedom have undergone a change of basis to the modes $|v_i\rangle$, it is no longer necessarily true that the variance due to fluctuations in the null signal $|{s}(\hat\theta)\rangle$ in a given mode are negligible compared to the noise fluctuations in that mode. To import the mechanics of the usual $\chi^2$ goodness-of-fit test to this reduced-dimension space, it is important to ensure that the \lcdm{} variation for each mode inside $S$ is negligible compared to the noise. We will present a method for selecting such modes in section \ref{sec:methods}.

Assuming we find a set of such $|v_i\rangle$s, our statistic is now given by the reduced-degree $\chi^2$ statistic
\begin{equation}
\label{eq: reduced chi2}
p_{\rm S} = \sum_{|v_i\rangle ,|v_j\rangle \in {\rm S}} \langle v_i | d - {s}(\hat\theta)\rangle N^{-1}_{ij} \langle d - {s}(\hat\theta) | v_j \rangle ,
\end{equation}
which under the null model should follow a $\chi^2$ distribution with $s$ degrees of freedom. Additionally, extreme values of $p_{\rm S}$ are plausibly explained by exotic cosmological fluctuations. It is worth noting that the relation in equation \ref{eq: signal-to-noise} has two methods of being satisfied: choosing either $|v_i\rangle$ that increase correlation with the expected signal, or choosing directions low in noise.  Binning, a widely used method of improving $\chi^2 $ sensitivity to the model, provides the latter improvement. In this case, the $|v_j\rangle$ in $S$ are a set of window functions of some width(s) $w_j$. As long as these widths of each bin are smaller than the natural correlation length of the expected signal, the degrees of freedom corresponding to the bins satisfy equation \ref{eq: signal-to-noise} through suppressing the right hand side of the inequality more than the left hand side.

The particular kind of extra-\lcdm{} fluctuations that reduced-degree $\chi^2$ is sensitive to is dependent on the kind of signals present in the prior distribution of fluctuations encapsulated in $\Sigma_\Xi$.
This is a general feature of a goodness-of-fit test that is often hidden in less explicit treatments. A reduced-degree $\chi^2$ test is very dependent on the choice of basis vectors that are included--it is more sensitive to signals that correlate well with the basis vectors kept, e.g. those $\Sigma_\Xi$ that satisfy equation \ref{eq: signal-to-noise}. 

%
%
\section{Optimal Linear Filtering}
\label{sec:methods}
We wish to find a set of $|v_i\rangle$ that appropriately tunes the degrees of freedom contributing to our $\chi^2$ statistic towards those with a high extra-\lcdm{} signal-to-noise ratio, e.g. those that satisfy equation \ref{eq: signal-to-noise}. To do this, we assume a model for an extra-\lcdm{} signal with data-space covariance $\Sigma_\Xi$. For a model with parameters $\xi_i$ with approximately linear responses over the plausible region of parameter space, this covariance is well approximated by
\begin{equation}
\Sigma_\Xi = \frac{\partial |{s_\Xi}(\boldmath{\xi_o})\rangle}{\partial \xi_i} \langle (\xi-\xi_o)_i (\xi-\xi_o)_j \rangle \frac{\partial \langle{s_\Xi} (\boldmath{\xi_o} )|}{\partial \xi_j},
\end{equation}
with $\langle (\xi-\xi_o)_i (\xi-\xi_o)_j \rangle$ the covariation of $\xi_i$ with $\xi_j$ in the prior $\mathcal{P}(\xi)$. In this expression we have assumed that $\xi_0$ is the mean of $\mathcal{P}(\xi)$ and that $\mathcal{P}(\xi)$ is centered at the \lcdm{} expectation (i.e. no exotic signal).
We also assume a measurement $|d \rangle$, which includes both \lcdm{} and extra-\lcdm{} signal:
\begin{equation}
|d\rangle = |s_\Lambda\rangle + |s_\Xi\rangle + |n\rangle,
\end{equation}
with noise $|n\rangle$ drawn from a model for the experimental noise $\mathcal{P}(|n\rangle)$ with covariance $\langle n_i n_j\rangle = N_{ij}$ and mean $\langle n_i \rangle=0$.

Our new degrees of freedom are given by
\begin{equation}
\label{eq: dof}
y_i = \langle v_i | d \rangle = \langle v_i | (|s_\Lambda\rangle + |s_\Xi\rangle + |n\rangle).
\end{equation}
We wish to minimize the contribution from $|n\rangle$ and $|s_\Lambda\rangle$ while maximizing the contribution from $|s_\Xi\rangle$. We define a ratio $\lambda$ for these degrees of freedom given by
\begin{equation}
\label{eq: signal to noise}
\lambda_i = \frac{E\left(\left| \langle v_i | s_\Xi \rangle \right|^2\right)}{E\left(\left|\langle v_i | \left(|n\rangle + \kappa |s_\Lambda \rangle\right)\right|^2\right)},
\end{equation}
where $\kappa \in \mathbb{R}$ is a relative weighting of \lcdm{} and noise fluctuations and $E()$ denotes averaging over draws from the priors on the exotic and null model parameters as well as from the noise model.
For reasons we give in section \ref{sec:kappa}, we take $\kappa \gg 1$.
We want the top $s$ filters $|v_1\rangle ,..., |v_s \rangle$ with highest $\lambda_1, ... ,\lambda_s$.
This is equivalent to requiring the filters $|v_i \rangle$ to satisfy the generalized eigenvalue relation
\begin{equation}
\label{eq: LDA modes}
\Sigma_\Xi |v_i \rangle = \lambda_i \left(\kappa^2 \Sigma_\Lambda + N\right) |v_i \rangle,
\end{equation}
with $\Sigma_\Lambda$ the signal space \lcdm{} prior covariance as detailed in section \ref{sec:kappa}.
Equation \ref{eq: LDA modes} can be solved for a set of $|v_i\rangle$ ordered in decreasing signal-to-noise (given by $\lambda_i$).

To construct a test, we simply choose the set $S$ which obeys equation \ref{eq: signal-to-noise} for some cutoff. For definitiveness in this paper, we choose the cutoff so that $\lambda_i \ge 0.05\lambda_0$, where $\lambda_0$ corresponds to the highest signal-to-noise mode. In principle, this choice can be guided \emph{a priori} (before performing the goodness-of-fit test) by examining the shape of the generalized eigenspectrum $\lambda_i$; however, in some cases (as for the model we analyze in section \ref{sec:p_of_k}) there is no natural break. However this cutoff is chosen, we then compute $p_S$ defined in equation \ref{eq: reduced chi2}, which follows a $\chi^2$ distribution with $s$ degrees of freedom under the null model. The test is then the PTE given by equation \ref{eq:PTE} with $m = s$. If the PTE for $p_S$ is small, we count this as evidence that the null model provides an inadequate description of the data.

\subsection{The role of the \lcdm{} signal in defining the $|v_i\rangle$}
\label{sec:kappa}

\begin{figure}{}
\includegraphics[width=0.5\textwidth]{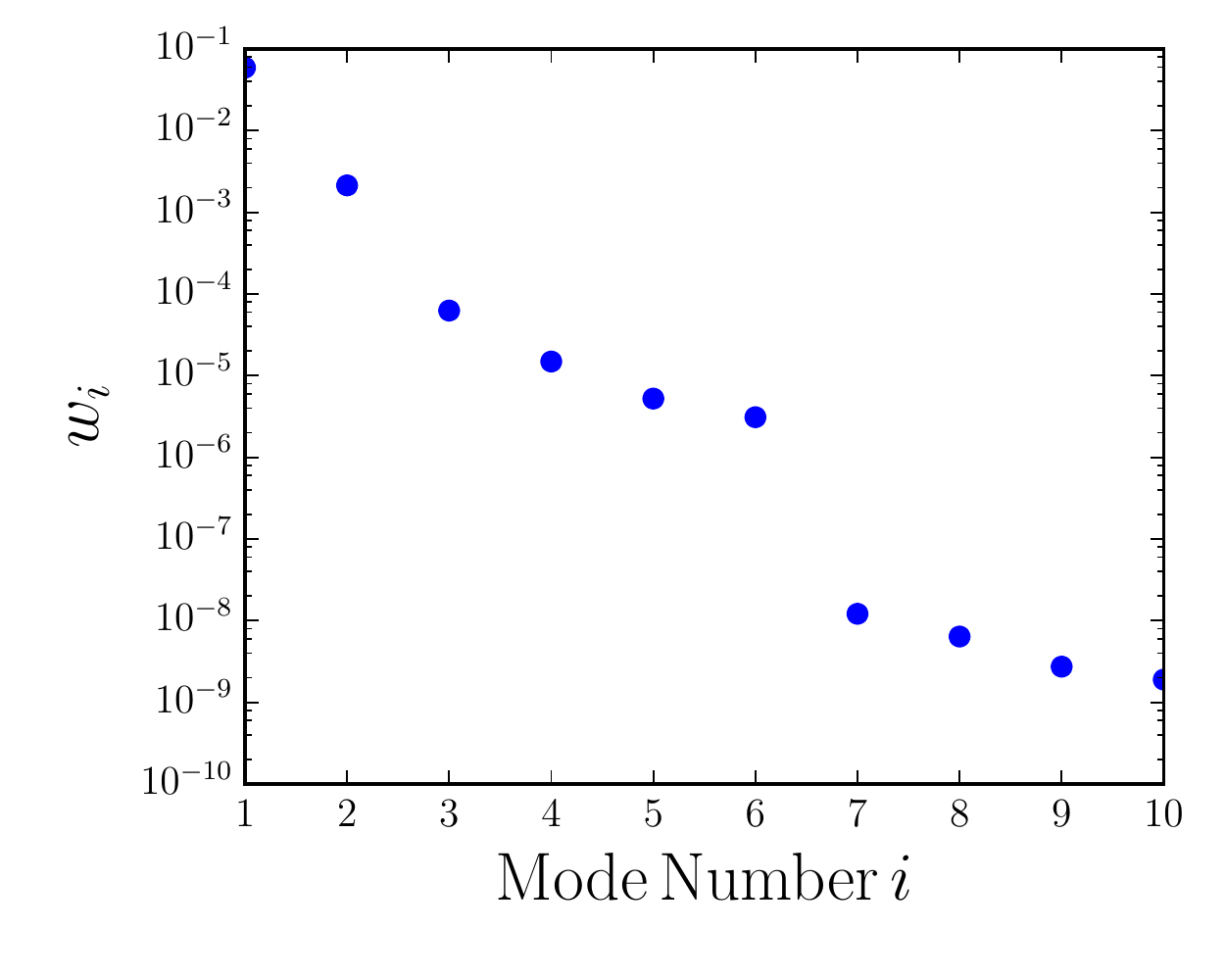}

\caption{\label{fig:lcdm eigenspectrum}
The largest eigenvalues of the \lcdm{} covariance matrix $\Sigma_\Lambda$, calculated using CLASS from a WMAP-derived prior. The first six modes contain all but $4.2\times 10^{-5} \%$ of the total power.}
\end{figure}

Equation \ref{eq: signal to noise} includes a term proportional to the \lcdm{} signal \ket{s_\Lambda} in the denominator next to the noise, with an as-yet undetermined constant $\kappa$. This term plays a multiplicity of crucial roles in the \ket{v_i} defined in equation \ref{eq: LDA modes}. Perhaps most transparently, since we are interested in finding an extra-\lcdm{} signal \ket{s_\Xi} in the data vector \ket{d}, the contribution to the variation from the \lcdm{} signal acts to obscure our signal of interest. In other words, it acts like noise. In addition, since we are seeking modes in the data with low noise, the requirement that the variation in \ket{d} be dominated by $N_{ij}$ is potentially violated if correlation with \lcdm{} fluctuations isn't likewise suppressed. In modes \ket{v_i} defined by a version of equation \ref{eq: signal to noise} lacking the term proportional to \ket{s_\Lambda}, the resulting degrees of freedom \ket{y_i} could have fluctuations in \lcdm{} signal that meet or exceed the noise fluctuations. The expected result of equation \ref{eq: reduced chi2} is then no longer distributed as a $\chi^2$ distribution under the null model. Finally, any degree of freedom \ket{y_i} correlated at all with \lcdm{} fluctuations \ket{s_\Lambda} has the potential to be activated by shifts in \lcdm{} parameters. Given the strong prior cosmologists should and do adopt towards this standard concordance model, this would introduce unwanted ambiguity in the interpretation of the resulting PTE.

\begin{figure}
\includegraphics[width=0.5\textwidth]{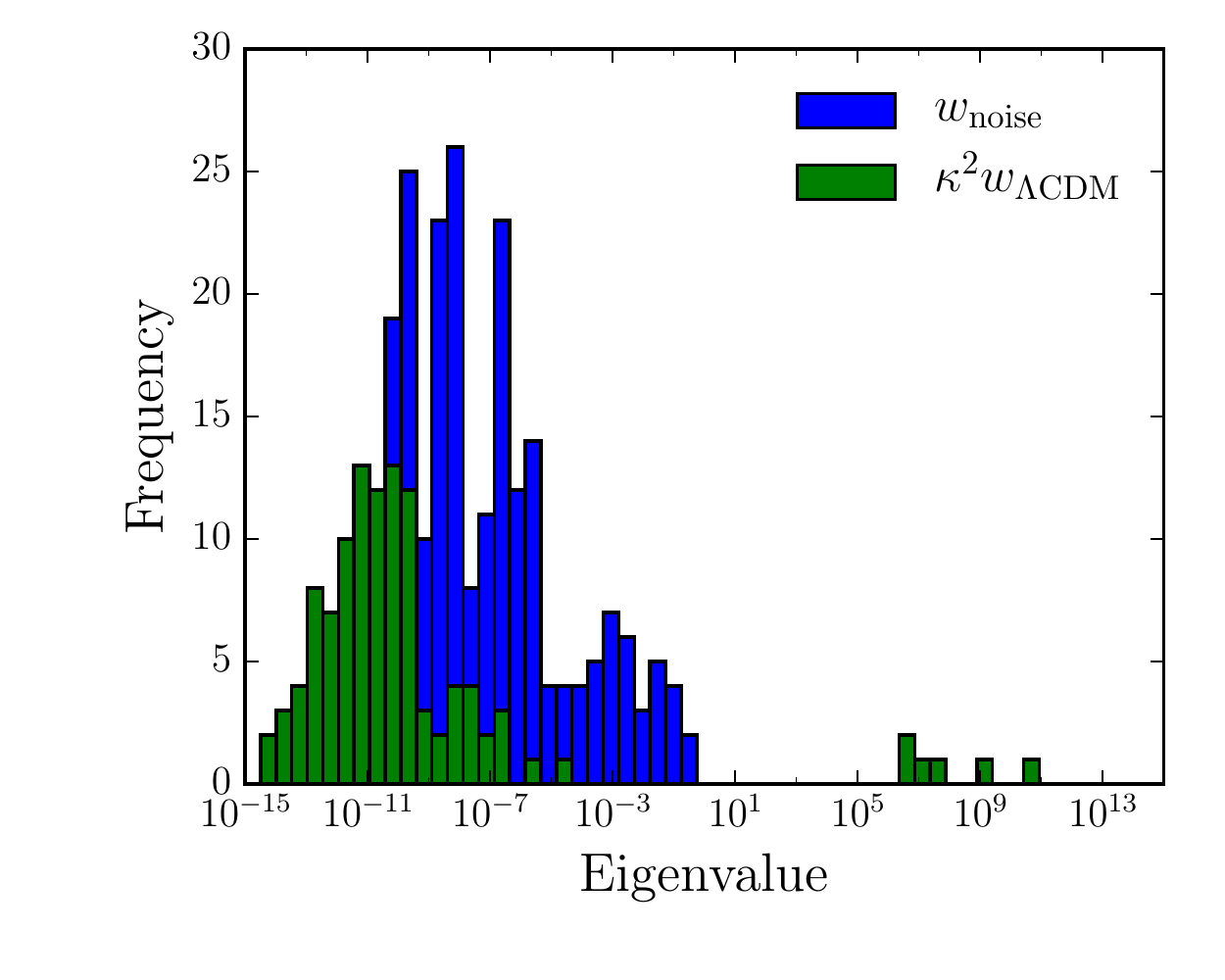}

\caption{\label{fig:lcdm top 6}The distribution of numerical eigenvalues of the two terms in the denominator of equation \ref{eq: signal-to-noise}, $N + \kappa^2 \Sigma_\Lambda$, with $\Sigma_\Lambda$ the reduced-rank covariance of section \ref{sec:kappa} and $\kappa$ set to $1 \times 10^6$.
Since we have kept only the top six modes in $\Sigma_\Lambda$, the eigenvalues of $\kappa^2 \Sigma_\Lambda$ smaller than $10^5 \mu K ^4$ are entirely due to numerical artifacts. Our choice of $\kappa$ is large enough that the top six eigenmodes of $\Sigma_\Lambda$ dominate the noise as we intend, but also small enough that these numerical artifacts do not compete with the highest several orders of magnitude of the eigenspectrum of $N$.}
\end{figure}

Fortunately, the \lcdm{} model is inherently low rank. In parameter space, the model is completely specified by a prior on the $6$ standard parameters $\theta = \left\{\log A, n_s, \tau, \omega_m, \omega_b, \text{ and } H_0 \right\}$, which we take from the WMAP9 posterior \citep{bennett_nine-year_2013}. When these parameters are mapped to angular power spectra with a Boltzmann code, in our case CLASS \citep{CLASS}\footnote{CLASS is available at http://class-code.net/}, this low rank structure leads to a steep drop in the eigenspectrum of the \lcdm{} prior covariance $\Sigma_\Lambda$ after the 6th eigenmode, as can be seen in Fig.~\ref{fig:lcdm eigenspectrum}. The vast majority of power in modes is in the span of a set of $6$ vectors that is stable over a wide range of priors on $\theta$. Therefore, in equation \ref{eq: signal to noise}, we approximate the covariance $\Sigma_\Lambda$ by the outer product of these $6$ vectors that span the subspace, and take $\kappa \gg 1$ to effectively project out this subspace. For the fiducial choice, we find $\kappa = 1 \times 10^6$ works well for de-projecting these $6$ modes without introducing any numerical instability into the solutions of equation \ref{eq: LDA modes}. Fig.~\ref{fig:lcdm top 6} shows the resulting eigenspectrum against the noise eigenspectrum it competes against in the denominator of equation \ref{eq: signal to noise}. So long as we do not make $\kappa$ large enough that the numerical instabilities become important or small enough that we fail to project out the $6$ top \lcdm{} modes, the value of $\kappa$ chosen does not significantly impact the analysis. Keeping only the top six eigenmodes allows us to avoid amplifying defects in the calculated covariance matrix that we  expect to arise from errors relating to the finite precision of the calculation of the power spectrum.

There is another use of the constant $\kappa$ that is not apparent in equation \ref{eq: LDA modes}. While above $\ell \simeq 30$ cosmic variance and shot noise are well-approximated by a Gaussian, non-Gaussian effects can be introduced by beam profiles and foregrounds that leave a characteristic pattern over multipoles. These are typically included in the noise by adopting some template $|T_i\rangle$ for each contribution, and adding a term $\alpha_i |T_i\rangle$ to the data model of equation \ref{eq:null model}. If these are counted as part of the noise, the effective noise matrix,
\begin{equation}
\tilde{N} = N_{\rm Gauss} + \sum_{i,j} \alpha_i \alpha_j |T_i\rangle \langle T_j|,
\end{equation}
is non-stationary with respect to variation in the parameters $\alpha_i$. However, if we instead include these degrees of freedom in the null model covariance $\Sigma_\Lambda$ of equation \ref{eq: signal to noise}, this dependence on $\alpha_i$ is projected out, and we recover the Gaussian condition. Since equation \ref{eq: LDA modes} only provides optimal modes if we have a Gaussian distribution for the noise model, this is an important consideration when considering data where non-Gaussian features contribute significantly to the total noise.

\section{Applications to Cosmology}
\label{sec:uses}

While the use of our statistic is, as far as we are aware, novel in cosmology, the type of filtering scheme described in section \ref{sec:methods} has a history. As early as the first year of publications based on the Cosmic Background Explorer (COBE) Satellite \citep{mather_measurement_1994}, various authors \citep{bond_signal--noise_1995,bunn_calculation_1996} used a variation of the $|v_i\rangle$ of equation \ref{eq: signal to noise} in pixel space (with $S_\Lambda = 0$ and $S_\Xi \rightarrow S_\Lambda$)
to compress CMB maps as an alternative to compression to the harmonic moments $a_{\ell m}$ or the power spectra $C_\ell$ at low multipoles. Like these standard compression techniques, the compression was computationally motivated. As computing power increased, these schemes were retired for the brute-force sampling of smoothed maps.

More recently, similar schemes to that of section \ref{sec:methods} based on Fisher matrix techniques have been used to investigate the power of CMB measurements to constrain the reionization history of the universe \citep{hu_model-independent_2003}, the mean-field inflaton potential $\phi(\eta)$ \citep{dvorkin_cmb_2010}, and energy deposition into the photon-baryon plasma from dark matter decay \citep{finkbeiner_searching_2012}. 
The methods in these papers differ most significantly from ours in that they are intended to select good parameter combinations to be varied in Monte Carlo Markov Chain (MCMC) methods and so select modes in parameter spaces instead of in $C_\ell$ space. 
As we argue in the introduction, this strategy leads to a narrower sensitivity to models, which in particular increases the dependence on the prior on extra-\lcdm{} signals. The choice advocated in this paper is to instead use a goodness-of-fit test statistic rather than directly fit for the presence of the features of the extra-\lcdm{} model in question, which reduces this prior sensitivity. Below we illustrate the general behavior of the statistic with a worked example.

\subsection{A worked example: primordial power spectrum generalization}
\label{sec:p_of_k}
A particularly interesting application is to a generalization of the primordial power spectrum.
As noted in e.g. \citet{planck_collaboration_planck_2014}, features in the inflaton potential lead directly to small-scale features in the power spectrum.

A concordance model of inflation does not yet exist \citep{martin_encyclopaedia_2013}, so the possibility exists that some small-scale feature, detectable through its effect on the power spectrum, can help narrow the zoo of inflationary models and provide insight to cosmological initial conditions and fundamental physics at the GUT scale.
As was previously done in \citet{planck_collaboration_planck_2014}, we can parameterize these effects on the power spectrum by considering a generalized spectrum given by
\begin{equation}
\label{general p_k}
\Delta^2_R(k) = \Delta^2_\Lambda(k) \left[1 + f(k)\right],
\end{equation}
where $f(k)$ is a parameterization of the fractional deviation from the \lcdm{} expectation $\Delta^2_\Lambda(k)$, given by
\begin{equation}
\Delta^2_\Lambda(k) = A_s \left(\frac{k}{k^*}\right)^{n_s - 1}.
\end{equation}
Here $A_s$ is the amplitude of the spectrum at $k = k^*$ and $n_s$ controls the scale dependence (with higher-order evolution such as running suppressed by slow roll).
In \citet{planck_collaboration_planck_2014}, the function $f(k)$ is represented as a cubic spline with knots separated by $\delta \ln k = 0.25$. The spline is fit with a regularization applied to both zero $f(k)$ outside the observable range of $k$ and to penalize high frequency oscillations. The strength of the regularization essentially introduces a minimum correlation scale in the space of allowed deviations $f(k)$ in the power spectrum. Here we do a similar analysis using the $Planck$ 2015 binned high-$\ell$ temperature data \citep{planck_collaboration_planck_2015}, but rather than fitting for the presence of a feature in the model, we instead use our goodness-of-fit statistic to search for evidence of features present in the data allowed by the generalized power spectrum model of equation \ref{general p_k} but not by \lcdm. We expect our results to be more general in the sense that they provide not only a search for deviations exactly parameterizable by equation \ref{general p_k}, but any feature that shares significant correlation with typical draws from equation \ref{general p_k}. More simply, through the use of goodness-of-fit rather than parameter estimation, we trade the ability to identify exactly the shape of $f(k)$ we are detecting for a sensitivity to a wider range of signals.

Our choice to work with the $Planck$ binned high-$\ell$ temperature data follows \citet{planck_collaboration_planck_2015-2} in binning the high-$\ell$ likelihood in bins sufficient to allow the folding of foreground uncertainty into the noise covariance. This choice eases both the computation and interpretation of the modes $|v_i\rangle$: $N$ in equation \ref{eq: LDA modes} represents the total noise covariance--experimental, foreground, and cosmic variance--present in the observation. In addition, since the cosmic variance at $\ell < 30$ is comparatively large, we discard the low-$\ell$ data where non-Gaussianities play a role. This justifies the implicit assumption in equation \ref{eq: LDA modes} that the structure of the noise can be taken to be Gaussian. 

\begin{figure}
\includegraphics[width=0.5\textwidth]{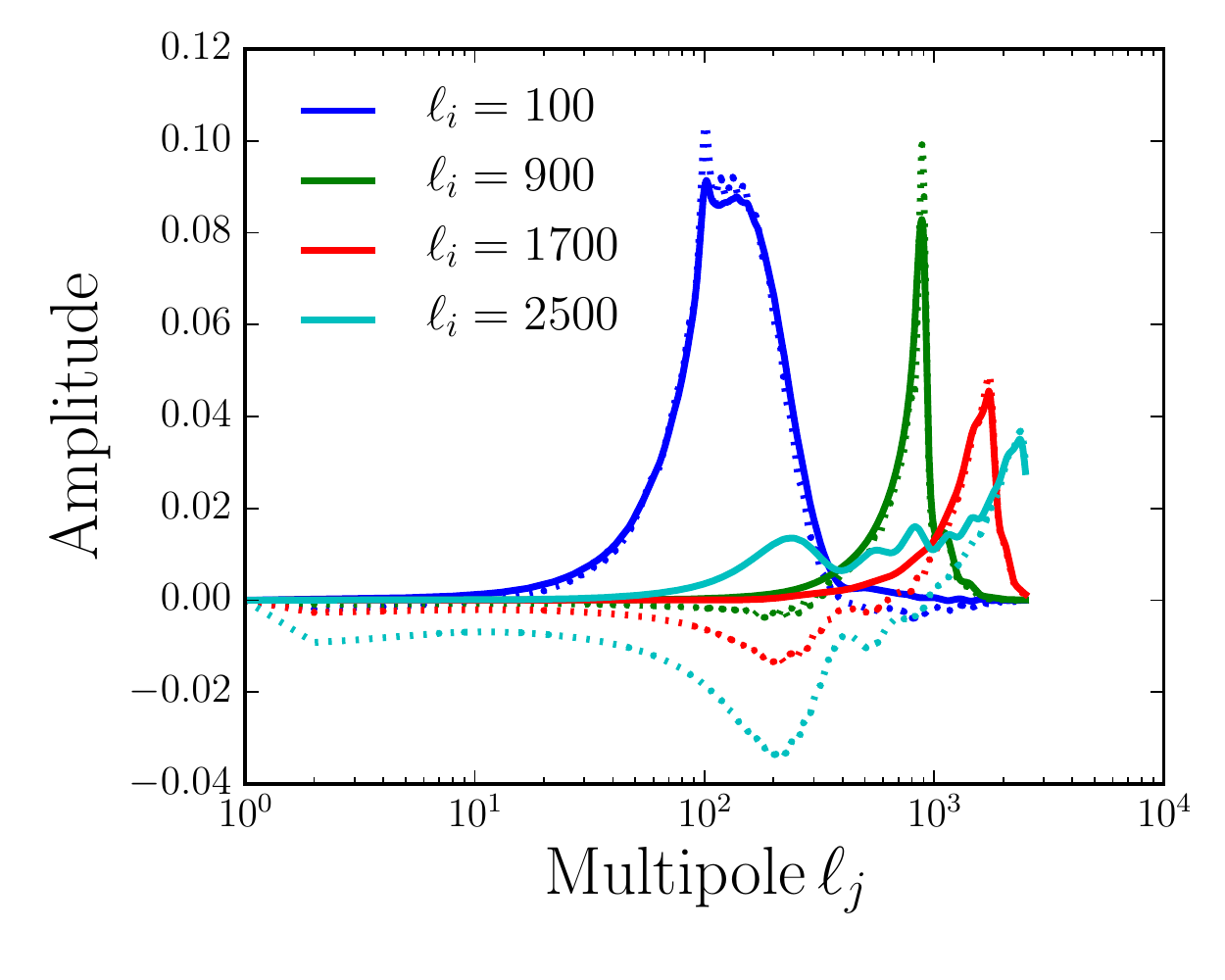}

\caption{\label{fig:exotic covariance}Columns from the covariance matrix ${\Sigma_\Xi}_{\ell_i\ell_j}$ for a diagonal prior on the $\xi_n$ of the Gaussian parameterization of equation \ref{p_k_gauss} (solid) and sinusoidal parameterization (dotted). Evident are the smoothing effects, which cause increasing correlation at increasing multipole $\ell_j$. The overall structure is broadly similar between the two parameterizations, reflecting the importance of the transfer function in the structure of the response to varying the $\xi_n$s.}
\end{figure}
\begin{figure}
 \includegraphics[width=0.5\textwidth]{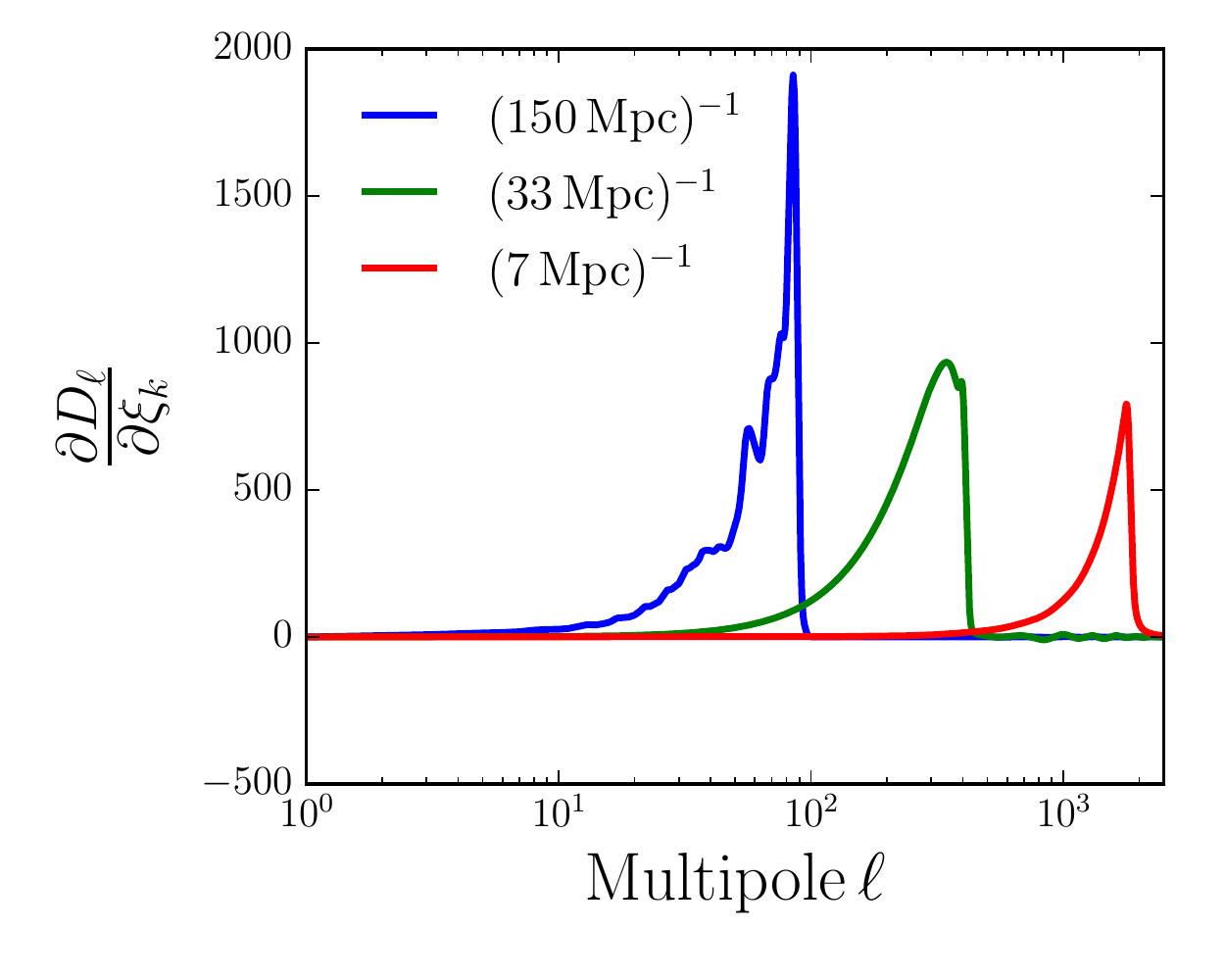}

 \caption{\label{fig:Gaussian_derivs}A characteristic subset of the $\frac{\partial |s_\Xi\rangle}{\partial \xi_n }$, where $|s_\Xi \rangle = \left[\ell (\ell + 1)/2\pi \right]
 C_\ell = D_\ell$, for the Gaussian parameterization of equation \ref{p_k_gauss}. Curves are labeled by the central wavenumber $k$ of the corresponding window $w_n$. The majority of the power put into the localized bumps $\Delta^2(k)$ translates to peaks in the $D_\ell$'s.}
\end{figure}
Though the features of the primordial power spectrum could be arbitrarily peaked, \citet{hu_principal_2004} show that when these features are transported to the observed power spectra, projection from the surface of last scattering and gravitational lensing smooth features below a characteristic width in $\Delta \ln k$. This provides a minimum frequency of variation in $f(k)$ observable in the data, and motivates us to parameterize $f(k)$ as a set of $\delta$-function like Gaussian bumps of characteristic width $\sigma$:
\begin{equation}
\label{p_k_gauss}
f(k) = \sum_n \xi_n \exp{\left(-\frac{ (\ln k - \ln k_n)^2}{2 \sigma^2}\right)},
\end{equation}
where, motivated by \citet{hu_principal_2004}, we take $\sigma = 0.25$ and choose $k_n$ spaced logarithmically by $\sigma$ in $\left[2 \times 10^{-4}, 8.2 \right] [{\rm Mpc}]^{-1}$. We assign parameters $\xi_n$ a prior covariance $\langle \xi_n \xi_m \rangle = \delta_{mn}$, and approximate the covariance $\Sigma_\Xi$ of the model on multipole space by approximating the response on the $C_\ell$s as linear, e.g.
\begin{equation}
\label{prior}
\Sigma_\Xi = \frac{\partial |s_\Xi\rangle}{\partial \xi_n } \langle \xi_n \xi_m \rangle \frac{\partial \langle s_\Xi|}{\partial \xi_m}.
\end{equation}
As for the calculation of $\Sigma_\Lambda$, these derivatives are calculated using CLASS \citep{CLASS}. To illustrate the effect of a different prior $\Sigma_\Xi$ on the $|v_i\rangle$ generated by equation \ref{eq: LDA modes}, we also investigate the case where the function $f(k)$ is instead parameterized by a sinusoidal basis,
\begin{equation}
\label{p_k_sines}
f(k) = \sum_n \left(\xi_{n,s} \sin \frac{2\pi n \ln (k)}{\lambda_{\rm max}} + \xi_{n,c} \cos \frac{2\pi n \ln (k)}{\lambda_{\rm max}}\right),
\end{equation}
where $k$ is measured in 1/Mpc. The wavelengths given by $\lambda_{\rm max}/n$ span the space $\left[0.05, 8\right]$ in order to cover the needed Fourier modes to approximate any arbitrary function with the maximum length size set by the observable range of the power spectrum. The minimum wavelength set to $.05$, is again motivated by \citet{hu_principal_2004}. Since these parameterizations cover a similar space of fluctuations $f(k)$ (in that any fluctuation generated by equation \ref{p_k_gauss} could be approximated by equation \ref{p_k_sines} for some choice of the parameters $\xi_{n}$), this re-parameterization effectively represents a change of prior $\langle \xi_n \xi_{n^\prime} \rangle$ in equation \ref{p_k_gauss}.

Fig.~\ref{fig:exotic covariance} shows columns of the correlation matrix $\Sigma_\Xi$ for selected multipoles $\ell_i$ under the two parameterizations given by equations \ref{p_k_sines} and \ref{p_k_gauss}. Despite the very different parameterizations represented by equations \ref{p_k_sines} and \ref{p_k_gauss}, the structures of the resulting covariance matrices are broadly similar to each other. This shows that the correlation scales are due primarily to projection effects, where scales of wavenumber $k$ project to larger scales through incidence on the surface of last scattering at acute angles.
This can be seen in the derivatives $\frac{\partial |s_\Xi\rangle}{\partial \xi_n }$ under the Gaussian parameterization shown in Fig.~\ref{fig:Gaussian_derivs}, where the correlation lengths in $\Sigma_\Xi$ are mirrored. The relative amplitudes of the responses are set by the shape of the fiducial model's angular power spectrum. 

\begin{figure}
\includegraphics[width=0.5\textwidth]{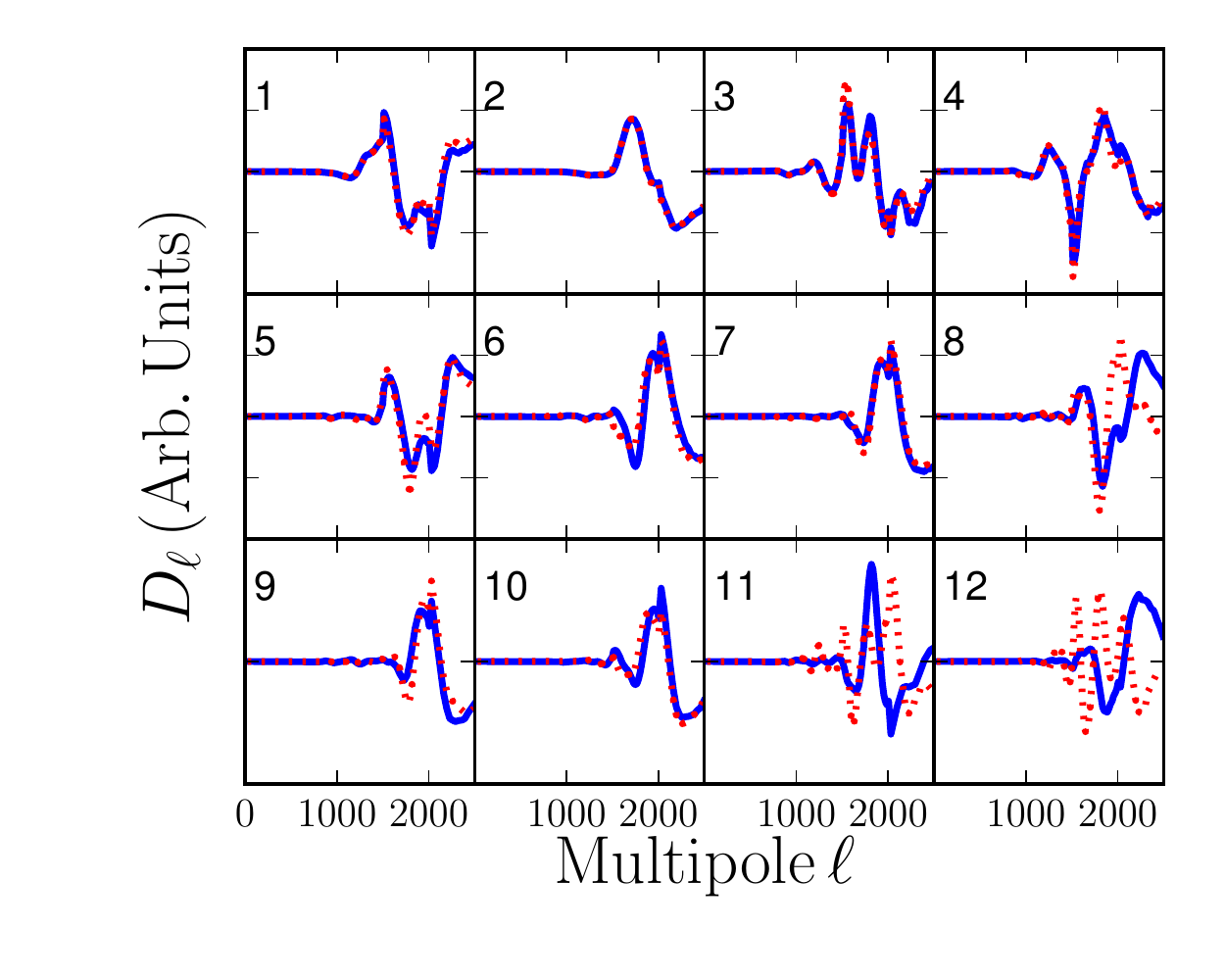}

\caption{\label{fig:LDA modes}
The modes $|v_i\rangle$ for the Gaussian perturbation model (blue, solid) and the sinusoidal perturbation model (red, dotted). The broad similarity of the filters is due to the dominance in the noise matrix $N$ in setting the modes in equation \ref{eq: LDA modes}. The lack of power at low multipoles is due to cosmic variance, and it is clear that most of the power is in the region with the smallest error bars from $Planck$, $\ell \in \left[\sim 1500, 1800 \right]$. The effect of projecting out \lcdm{} directions is to project out variations in well-measured angular scales, most notably the sound horizon $\theta_s$.}
\end{figure}

\begin{figure}
\includegraphics[width=0.5\textwidth]{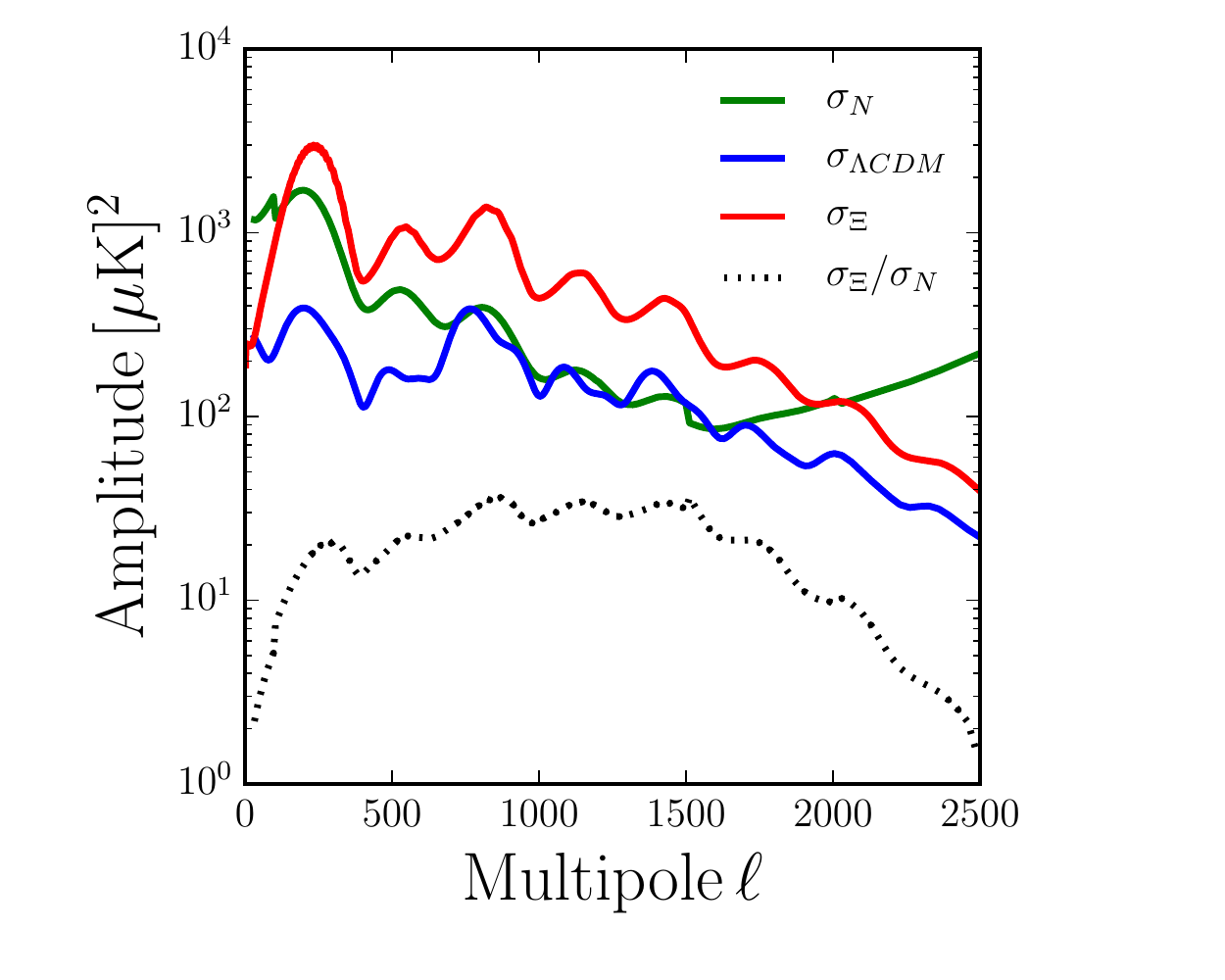}
\caption{\label{fig:noise v exotic}
The square root of the diagonal elements of the noise covariance $N$, the \lcdm{} signal covariance $\Sigma_\Lambda$, and the linearized multipole-space covariance $\Sigma_\Xi$ for the Gaussian perturbation model under the prior given by equation \ref{prior} as well as their ratios. Plots are in units of $D_\ell = \left[\ell (\ell + 1)/2\pi\right]C_\ell$. The variance in each model for $\Sigma_\Xi$ contains an arbitrary offset set by the scale of the prior on the $\xi_n$'s. Since the exotic fluctuations are drawn from primordial spectra perturbations with uniform probability across wavenumber $k$, the subsequent transfer of power creates a diagonal structure similar to a CMB realization under \lcdm{} with a flat primordial spectrum ($n_s = 1$). Off-diagonal correlations are of course quite different, and are shown in Fig.~\ref{fig:exotic covariance}.  The sensitivity of the $Planck$ experiment to the multipole region around $\ell \simeq 1600$ which our filters are tuned to is clearly visible. Note that the filters in Fig.~\ref{fig:LDA modes} do not emphasize the higher $\sigma_\Xi$ values at lower multipoles where there is also more noise.}
\end{figure}

The resulting filters for both the sinusoidal and Gaussian parameterizations are shown in Fig.~\ref{fig:LDA modes}.
As one would expect from the broad similarities of the covariance structure for both parameterizations, the highest signal-to-noise vectors are very similar, with more variation shown in vectors $5$ through $9$ to accommodate the small scale differences due to the change in parameterization. Note that the structure of the modes reflects the $\ell$ range that has both the highest signal to noise and the smallest \lcdm{} signal (Fig.~\ref{fig:noise v exotic}). 


\begin{figure}
\includegraphics[width=0.5\textwidth]{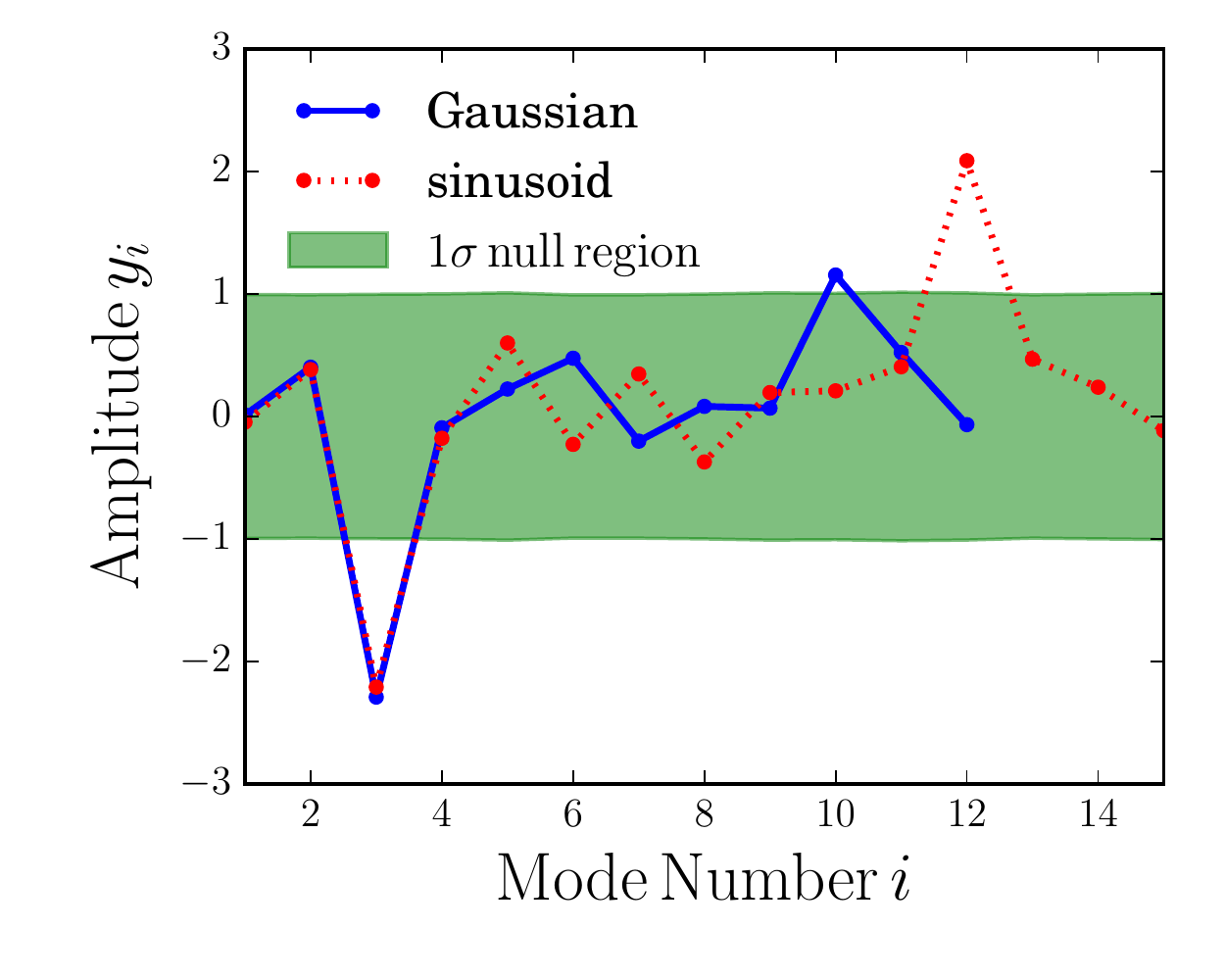}
\caption{\label{fig:filtered_data} The measured degrees of freedom $\langle d | v_i \rangle $ under the truncated basis $|v_i\rangle$ chosen by equation \ref{eq: LDA modes} under the exotic covariance $\Sigma_\Xi$ of the Gaussian parameterization of equation \ref{p_k_gauss} and the diagonal prior of equation \ref{prior}. The normalization of the $|v_i \rangle$ are chosen so that a draw from the \lcdm-deprojected noise covariance $N + \kappa^2 \Sigma_\Lambda$ has expected value one. The green region represents the one $\sigma$ region for $10000$ simulated observations under the null \lcdm{} model, where the simulations use the full-rank \lcdm{} covariance $\Sigma_\Lambda$ with prior given by the WMAP posterior. That this region is consistent with one across all degrees of freedom demonstrates the insensitivity of the $|v_i\rangle$ to higher-order contributions of \lcdm{} signal $|s\rangle$.} 
\end{figure}

\begin{figure}
\includegraphics[width=0.5\textwidth]{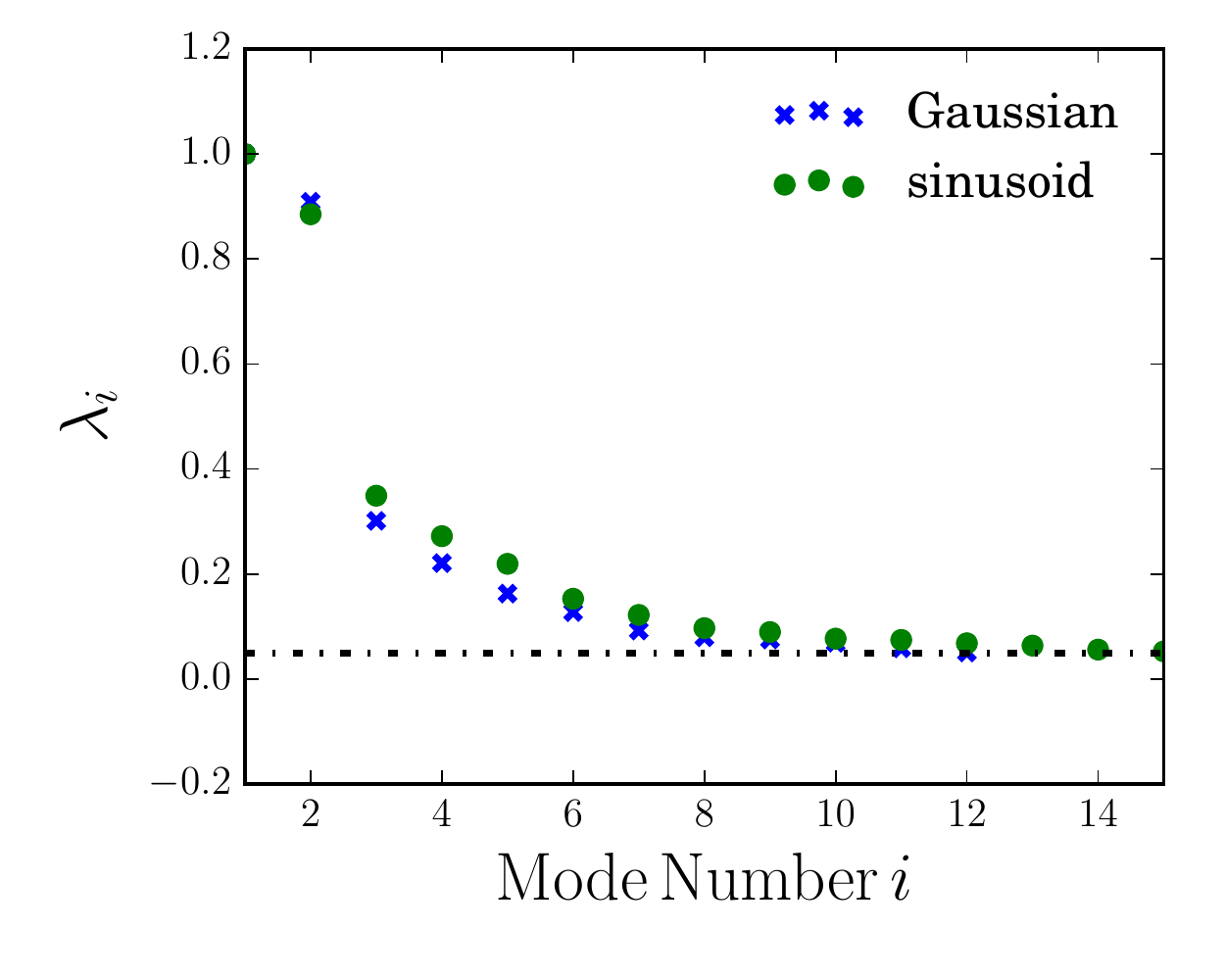}

\caption{\label{fig:LDA_spectrum}The top signal-to-noise ratios $\lambda_i$ (satisfying $\lambda_i \ge 0.05 \lambda_{\rm max}$) of the basis vectors found in equation \ref{eq: signal-to-noise} with the exotic signal given in equation \ref{p_k_gauss}, with the diagonal prior given by equation \ref{prior}. The prior is scaled until $\lambda_{\rm max} = 1$, and the cutoff is denoted with the dot-dash line. The Gaussian parameterization is shown as blue x's and the sinusoidal parameterization is shown as green circles. As noted in section \ref{sec:methods}, there is no break here where there is a significant jump in the order of magnitude of the $\lambda_i$'s. The largest break that is visible in the plot is only a jump by a factor of roughly 3, not a factor of around $10^2$ as in the \lcdm{} eigenspectrum.}
\end{figure}

Examining Fig.~\ref{fig:filtered_data}, we see that for both parameterizations of $f(x)$, the most excited mode in the data is the third, with an amplitude distinguishable from the \lcdm{} expectation at around $2.5\sigma$. For the sinusoidal parameterization the twelfth mode is of comparable significance, though since the signal-to-noise eigenspectrum (Fig.~\ref{fig:LDA_spectrum}) is less dominated in the sinusoidal parameterization by the first couple of modes the significance of having this additional excursion is offset by the inclusion of more degrees of freedom in the $\chi^2$ test given our chosen cutoff of $\lambda_i \ge 0.05 \lambda_{\rm max}$. 

We can also see the redistribution of power among the modes due to the re-parameterization in Fig.~\ref{fig:filtered_data}; while the amplitudes of the data projected onto the modes from each parameterization are somewhat different, the significance of the test is broadly unaffected due to the similarity of the filters derived with each parameterization (Fig.~\ref{fig:LDA modes}). The excursions for most modes lie within the $1 \sigma$ expectation, and the Gaussian and sinusoidal parameterizations lead to PTE$ = 0.83$ and PTE$ = 0.78$, respectively, which we interpret as a lack of evidence for the presence of primordial perturbations $f(k)$.

It is worth noting again that even if the PTE were lower (say PTE$=0.0005$), this is also not direct evidence for the presence of perturbations given by equation \ref{general p_k}. A goodness-of-fit test is not a direct comparison of two models, but rather a test of the null hypothesis. The correct interpretation of a small PTE is general evidence for some extra-\lcdm{} fluctuations or systematic error in the data--the trade-off for the generality of the test compared to parameter fitting is that a positive result doesn't point in a specific direction. A feature in the data that correlates well with one of the filters $|v_i\rangle$ over just a small range in multipoles $\ell$ can lead to a significant excursion even if outside of this range the data correlate with the model poorly--in other words, a `detection' may point to a feature that isn't even reproducible in the context of the model in question.

\subsection{Including polarization information}
\begin{figure}
\includegraphics[width=0.5\textwidth]{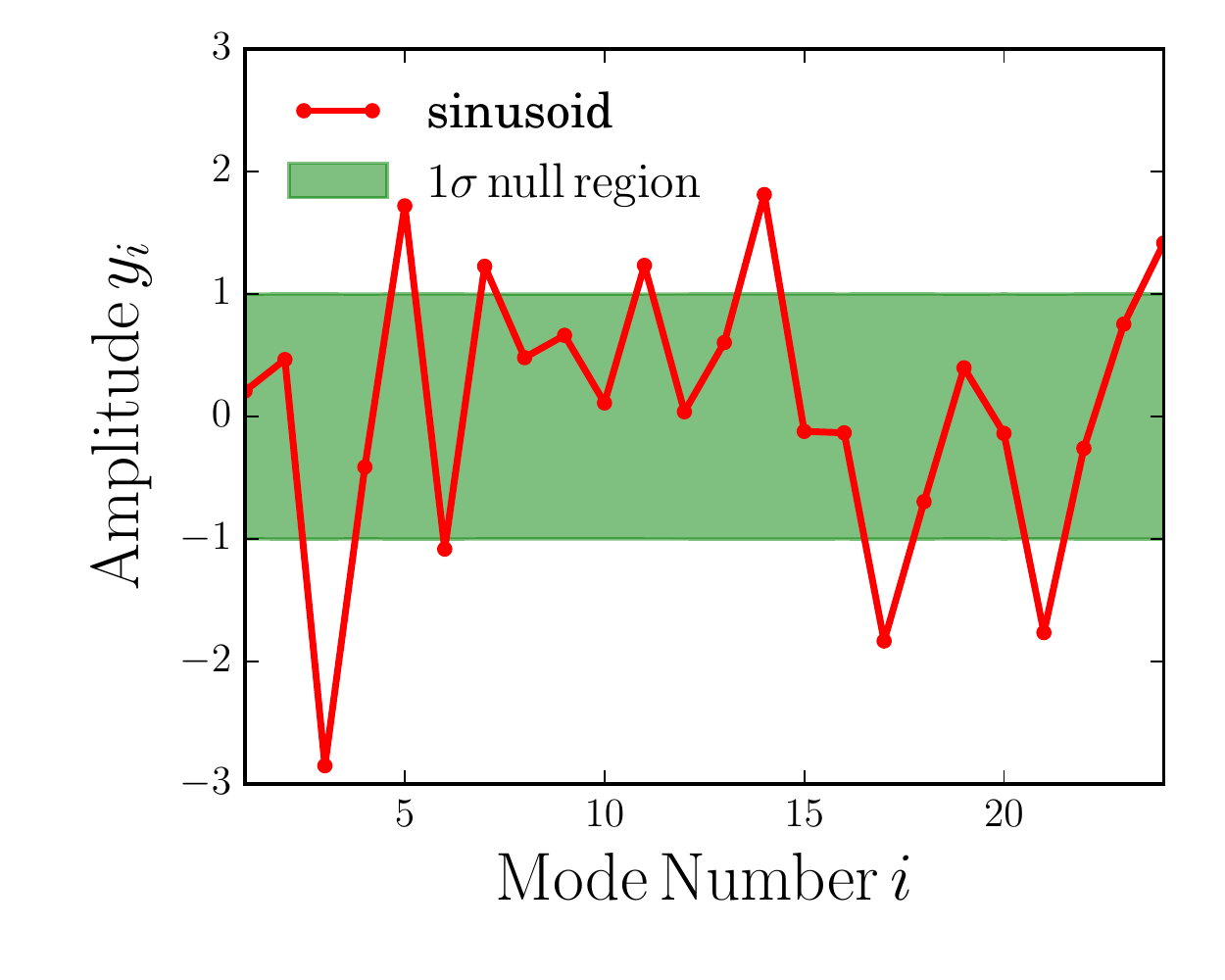}
\caption{\label{fig:filtered_data TE} Same as Fig.~\ref{fig:filtered_data} for the sinusoidal parameterization, but including both temperature and polarization data from $Planck$. The deviation in the third degree of freedom $y_3$ from the null expectation is not significant over the entire set of $25$ degrees of freedom that satisfy $\lambda_i > 0.05 \lambda_0$. The $\sim 3 \sigma$ signal one gets if one ignores a look-elsewhere effect can potentially be explained by observed temperature polarization leakage in the most recently release $Planck$ polarization information.
}
\end{figure}

The above subsection restricted our search for $P(k)$ deviations to the CMB temperature power spectrum. 
This was done for both pedagogical simplicity, and because the most recent polarization data from $Planck$ has known systematics effects that have not yet been fully characterized \citep{planck_collaboration_planck_2015-2}. 
Nevertheless, an extension to polarization is straightforward through extending the sinusoidal parameterization of equation \ref{p_k_sines} to include information in the E-mode autospectrum and the TE cross-spectrum. 
Including this polarization information leads to a PTE of $0.19$ over the $25$ filtered degrees of freedom shown in Fig.~\ref{fig:filtered_data TE}.

\begin{figure}
\includegraphics[width=0.5\textwidth]{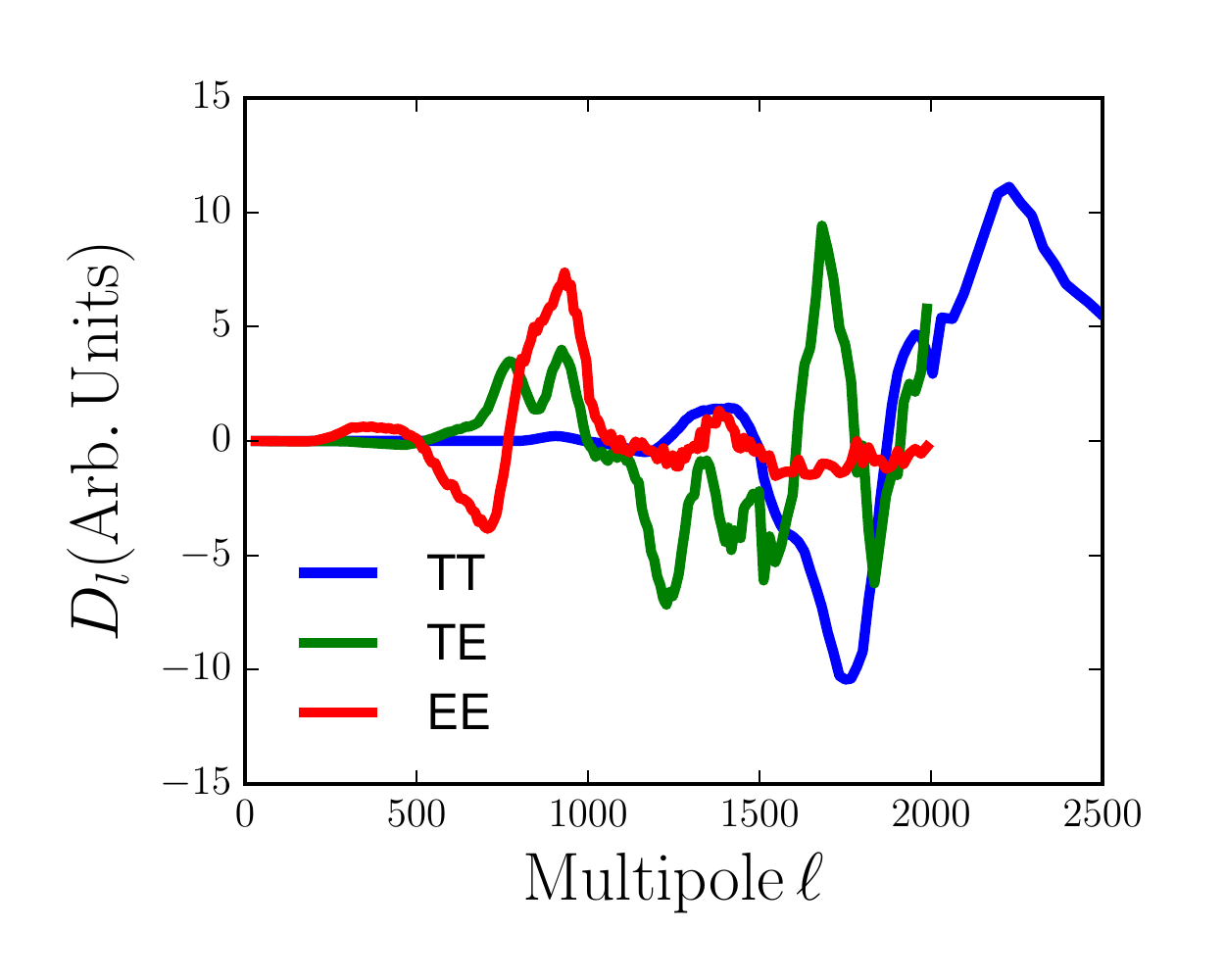}
\caption{\label{fig:filter TE}
The shape of the filter corresponding to the filtered degree of freedom with maximum departure $|v_3\rangle$ when including E-mode polarization auto spectra and cross spectra with temperature. The filter results in a direction that is in roughly $3\sigma$ tension with the null \lcdm{} model, with most of this departure due to multipoles around $\ell \simeq 1000$ in the TE spectrum (green). The general shape of the filter in this region is similar to known TE systematics from temperature-polarization leakage.
}
\end{figure}

\begin{figure}
\includegraphics[width=0.5\textwidth]{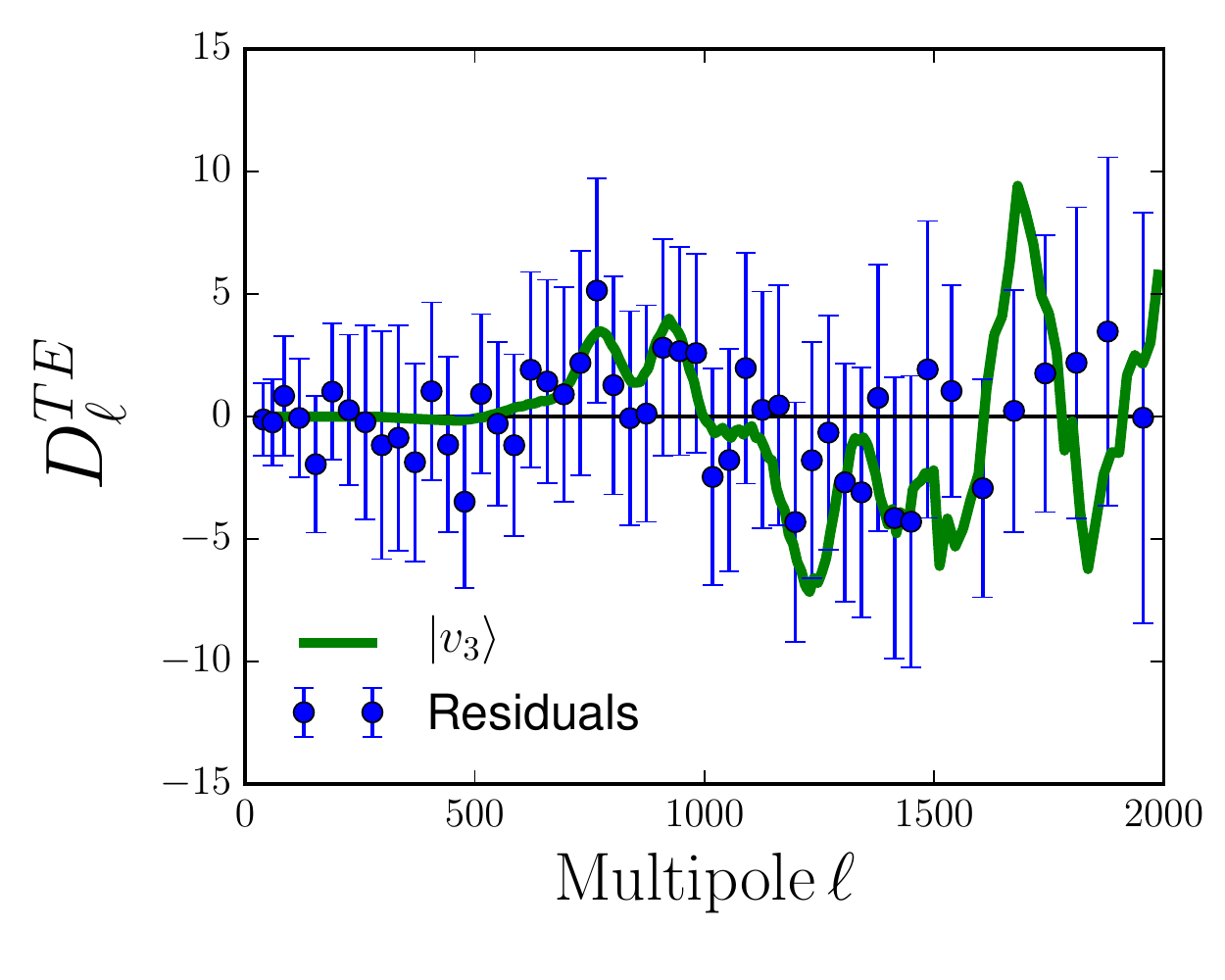}
\caption{\label{fig:TEresid}
The residuals from the best fit \lcdm{} power spectrum, re-binned to reduce the number of plotted data points for visual clarity. The pattern in these residuals, which can be largely attributed to known TE systematics from temperature-polarization leakage, gives us the departure from the \lcdm{} expectation on $|v_3\rangle$, which is also shown. Note that our filters, including this $|v_3\rangle$, are generated before and not determined by the measured data. The amplitude of $|v_3\rangle$ was chosen to roughly correspond to the scale of the residuals.
}
\end{figure}

Of potential interest is the $\sim 3 \sigma$ departure in the amplitude of the third mode $y_3$, shown in Fig.~\ref{fig:filter TE}. The value of $y_3$ is determined through equation \ref{eq: dof}; the dominant contribution to the dot product is in the intermediate multipoles $\ell \in \left[600, 1200\right]$ of the TE cross-spectrum. The residuals from the $Planck$ best fit $C_\ell$'s for this region are shown in Fig.~\ref{fig:TEresid}, where we can see what pattern in the data is being detected with this mode.

This multipole range matches where we expect to see the effects of TT leakage on the TE cross-spectrum as reported in \citet{planck_collaboration_planck_2015-2}. 
In addition, the shape of the filter in this range is broadly similar in shape to the filter shown in Fig. $21$ of \citet{planck_collaboration_planck_2015-2}. 
It is therefore plausible that the activation of $|v_3\rangle$ when extending to polarization is due to this known systematic effect, though a rigorous analysis of this effect must await the public release of the data products used in \citet{planck_collaboration_planck_2015-2}.

\section{Summary and Conclusion}
\label{sec:conclusion}
We have presented a general method of tuning goodness-of-fit tests to locate deviations from a standard model that are motivated by specific model extensions. This method falls between a $\chi^2$ test on the full data degrees of freedom and parameter estimation in that it is particularly sensitive to models with similar features to the model it is tuned towards, without the `extreme' tuning inherent in considering a fit to a particular model. The increased generality of the test compared to traditional model comparison methods, such as fitting for a laundry list of extensions of a standard model, makes this method a powerful tool for confronting new and existing data. 

We note the method is particularly useful in cases where one is interested in a family of models, not easily parameterized but with similar features. This can be a family of model extensions, but the method also holds promise as a test for specific classes of contaminants in the data.

These features make the method a very useful tool in cosmology where we have a standard model (\lcdm) with many possible extensions and high dimensional datasets that necessarily have contaminants from foreground contributions. 

As a demonstration of our method, we have presented such tuned goodness-of-fit tests to look for deviations from \lcdm{} that are similar to those sourced by generalizations of the primordial power spectrum in the $Planck$ CMB temperature data and then both temperature and polarization data. Since we do not find significant evidence for any such deviations, we conclude that there is not sufficient motivation to carry out further investigations into possible signals from, for example, inflation that are purely encoded in the scalar power spectrum with this data set.

We also note that in this example our test found a $~3\sigma$ departure on a mode which looks like it might be capturing some of the TT leakage on the TE cross-spectrum. Taken in the context of our test this departure on only one mode is not statistically significant, but the fact that a mode chosen based on perturbations to the primordial power spectrum can capture such leakage is an example of how our test fits between model fitting and a $\chi^2$ test on the data degrees of freedom in terms of the generality of the test. This feature was not tuned for and was definitely not detected as strongly as fitting for its presence would have, but it \emph{did} show up where it would not have with either a full $\chi^2$ test or directly fitting for power spectrum perturbations.

Finally, we reiterate that, as with any goodness-of-fit test, a failure of the null hypothesis does not count as evidence for any other model, even the model the test was tuned towards. If we find such a failure we are motivated to investigate further, perhaps by investing the effort and computational resources to carry out a sequence of specific model comparison tests.

\section*{Acknowledgments}
A.A. was supported in part by DOE Grant DE-SC-0009999. E.A. was supported in part by NSF CAREER grant DMS-1252795

\bibliographystyle{mnras}
\bibliography{LDA_merged}

\end{document}